# Multi-atom quasiparticle scattering interference for superconductor energy-gap symmetry determination

Rahul Sharma[1,2,§], Andreas Kreisel[3,§], Miguel Antonio Sulangi[§,4], Jakob Böker[5],
Andrey Kostin[1], Milan P. Allan[6], H. Eisaki[7], Anna E. Böhmer[8,9], Paul C. Canfield[9,10],
Ilya Eremin[5,11], J.C. Séamus Davis*[1,12,13,14], P.J. Hirschfeld[4], and Peter O. Sprau[1,15]

1. Department of Physics, Cornell University, Ithaca NY 14850, USA.
2. Maryland Quantum Materials Center, Department of Physics, University of Maryland, College Park, MD 20742, USA.
3. Institut für Theoretische Physik, Universität Leipzig, Brüderstr. 16, Leipzig 04103, Germany.
4. Department of Physics, University of Florida, 2001 Museum Rd, Gainesville, FL 32611, USA.
5. Institut für Theoretische Physik, Ruhr-Universität Bochum, D-44801 Bochum, Germany.
6. Leiden Institute of Physics, Leiden University, Niels Bohrweg 2, 2333CA Leiden, The Netherlands.
7. National Inst. of Advanced Industrial Science and Tech., Tsukuba, Ibaraki 305-8568, Japan.
8. Institute for Quantum Materials and Technologies, Karlsruhe Institute of Technology (KIT), Karlsruhe 76133, Germany.
9. Ames Laboratory, U.S. Department of Energy, Ames, Iowa 50011, USA.
10. Department of Physics, Iowa State University, Ames, Iowa 50011, USA.
11. Institute of Physics, Kazan Federal University, Kazan 420008, Russian Federation.
12. Department of Physics, University College Cork, Cork T12 R5C, Ireland.
13. Max-Planck Institute for Chemical Physics of Solids, D-01187 Dresden, Germany.
14. Clarendon Laboratory, University of Oxford, Oxford, OX1 3PU, UK.
15. Advanced Development Center, ASML, Wilton, CT 06897, USA.
§ These authors contributed equally to this project.

*corresponding author: jcseamusdavis@gmail.com

**Complete theoretical understanding of the most complex superconductors requires a detailed knowledge of the symmetry of the superconducting energy-gap $\Delta_k^\alpha$, for all momenta k on the Fermi surface of every band $\alpha$. While there are a variety of techniques for determining $|\Delta_k^\alpha|$, no general method existed to measure the signed values of $\Delta_k^\alpha$. Recently, however, a technique based on phase-resolved visualization of superconducting quasiparticle interference (QPI) patterns centered on a single non-magnetic impurity atom, was introduced. In principle, energy-resolved and phase-resolved Fourier analysis of these images identifies wavevectors connecting all k-space regions where $\Delta_k^\alpha$ has the same or opposite sign. But use of a single isolated impurity atom, from whose precise location the spatial phase of the scattering interference pattern must be measured is technically difficult. Here we introduce a generalization of this approach for use with multiple impurity atoms, and**

**demonstrate its validity by comparing the $\Delta_\mathbf{k}^\alpha$ it generates to the $\Delta_\mathbf{k}^\alpha$ determined from single-atom scattering in FeSe where $s_\pm$ type sign-changing energy-gap is established. Finally, to exemplify utility, we use the multi-atom technique on LiFeAs and find scattering interference between the hole-like and electron-like pockets as predicted for $\Delta_\mathbf{k}^\alpha$ of opposite sign.**

Keywords: Superconductor Order-Parameter Symmetry Determination, Multi-Atom Quasiparticle Scattering Interference

**INTRODUCTION**

The macroscopic quantum condensate of electron pairs in a superconductor is represented by its order-parameter $\Delta_\mathbf{k}^\alpha \propto \langle c_\mathbf{k}^{\alpha\dagger} c_{-\mathbf{k}}^{\alpha\dagger} \rangle$ where $c_\mathbf{k}^{\alpha\dagger}$ is the creation operator for an electron with momentum **k** on band $\alpha$. But electron pair formation can occur through a wide variety of different mechanisms and in states with many possible symmetries[1]. Thus, it is the symmetry properties of $\Delta_\mathbf{k}^\alpha$ that are critical for identification of the Cooper pairing mechanism[1] and, moreover, for understanding the macroscopic phenomenology[1]. While macroscopic techniques can reveal energy-gap symmetry for single-band systems[2,3], no general technique existed to determine the relative signs of $\Delta_\mathbf{k}^\alpha$ and $\Delta_{\mathbf{k}'}^\beta$ between $\mathbf{k}_\alpha$ and $\mathbf{k}_\beta$ for all Fermi surface momenta in an arbitrary superconductor.

In 2015 a conceptually simple and powerful technique for determining $\Delta_\mathbf{k}^\alpha$ symmetry was introduced[4], by Hirschfeld, Eremin, Altenfeld and Mazin (HAEM). It is based on interference of weakly scattered quasiparticles at a single, non-magnetic, impurity atom. Given a superconductor Hamiltonian

$$\mathcal{H}_k = \begin{pmatrix} H_\mathbf{k} & \Delta_\mathbf{k} \\ \Delta_\mathbf{k}^\dagger & -H_{-\mathbf{k}}^T \end{pmatrix}, \tag{1}$$

where $H_\mathbf{k}$ is the normal-state Hamiltonian and $\Delta_\mathbf{k}$ the superconducting energy gap, a non-magnetic impurity atom is modeled as a weak point-like potential scatterer, with Hamiltonian $H_\text{imp} = V_\text{imp} c_\mathbf{r}^\dagger c_\mathbf{r}$ centered at the origin of coordinates $\mathbf{r} = 0$. Effects of scattering are then represented by a T-matrix derived from the local Green's function

$G_0(E) = \sum_\mathbf{k} G_\mathbf{k}^0(E)$ where $G_\mathbf{k}^0(E) = (E + i0^+ - \mathcal{H}_\mathbf{k})^{-1}$. When the impurity potential is constant in **k**-space, the Green's function becomes $G_{\mathbf{k},\mathbf{k}'}(E) = G_{\mathbf{k},\mathbf{k}'}^0(E) + G_\mathbf{k}^0(E)T(E)G_{\mathbf{k}'}^0(E)$, with the T-matrix given by $T(E) = [1 - V_{\text{imp}}G_0(E)]^{-1}V_{\text{imp}}$. From $G_{\mathbf{k},\mathbf{k}'}(E)$, the perturbations to the local density-of-states $\delta N(\mathbf{r}, E)$ are predicted surrounding the impurity atom, and its Fourier transform can be determined directly from $\Delta_\mathbf{k}$ as

$$\delta N(\mathbf{q}, E) = -\frac{1}{\pi} \text{Im} \left[ \sum_\mathbf{k} G_\mathbf{k}^0(E) T(\omega) G_{\mathbf{k}+\mathbf{q}}^0(E) \right]_{11} \tag{2}$$

which is a purely real quantity because, in the theoretical calculation, the single impurity is exactly at the origin of coordinates. The authors of Ref. 4 realized that the particle-hole symmetry of Eq.(2) for scattering interference wavevector $\mathbf{q} = \mathbf{k}_f^\beta - \mathbf{k}_i^\alpha$, depends on the relative sign of the energy-gaps $\Delta_{\mathbf{k}_i}^\alpha$ and $\Delta_{\mathbf{k}_f}^\beta$ at these two momenta. Consequently, the experimentally accessible energy-antisymmetrized function $\rho^-(\mathbf{q}, E)$ of phase-resolved Bogoliubov scattering interference amplitudes

$$\rho^-(\mathbf{q}, E) \equiv \text{Re}\{\delta N(\mathbf{q}, +E) - \delta N(\mathbf{q}, -E)\} \tag{3}$$

can be used to determine the relative sign of the superconducting energy-gaps connected by $\mathbf{q} = \mathbf{k}_f^\beta - \mathbf{k}_i^\alpha$. In the simplest case with two isotropic gaps $\Delta^\alpha$ and $\Delta^\beta$ on distinct bands, it was demonstrated that

$$\rho^-(\mathbf{q}, E) \propto \text{Im} \left[ (E_+^2 - \Delta^\alpha \Delta^\beta) / \sqrt{E_+^2 - (\Delta^\alpha)^2} \sqrt{E_+^2 - (\Delta^\beta)^2} \right] \tag{4}$$

where $E_+ = E + i0^+$, so that the functional form of $\rho^-(\mathbf{q}, E)$ is very different when the product $\Delta^\alpha \Delta^\beta$ is positive or negative. An elementary implication of Eq.(4) is that, when order parameter has opposite signs on the two bands so that $\Delta^\alpha \Delta^\beta < 0$, $\rho^-(\mathbf{q}, E)$ does not change sign and exhibits pronounced maxima or minima near $E \approx \Delta^{\alpha,\beta}$ whereas if the order parameter has the same sign so that $\Delta^\alpha \Delta^\beta > 0$, $\rho^-(\mathbf{q}, E)$ exhibits weak maxima or minima near $E \approx \Delta^{\alpha,\beta}$ with a sign of change in between. More generally, especially with multiple bands and anisotropic gaps, HAEM requires that $\rho^-(\mathbf{q}, E)$ be predicted in detail for a specific $H_\mathbf{k}$ and $\Delta_\mathbf{k}$ in Eq. (1) and then compared with quasiparticle interference imaging[5] in which the STM differential electron tunneling conductance, $g(\mathbf{r}, E) \propto \delta N(\mathbf{r}, E)$ is visualized.

This single-atom phase-resolved HAEM method has been established experimentally[6,7]. For example, in the case of the multiband $s_\pm$ superconductor FeSe, the complete energy and wavevector dependence of $\rho^-(\mathbf{q}, E)$ was used to determine that the **k**-space structure including relative sign of $\Delta_\mathbf{k}^\alpha$ and $\Delta_\mathbf{k}^\beta$, for all $\mathbf{k}_\alpha$ and all $\mathbf{k}_\beta$ on two different bands. But this result required that the impurity atom be highly isolated from other impurities and centered precisely at the origin of coordinates, with respect to which the Re$\delta N(\mathbf{q}, E)$ of Eq.(3) is then properly defined. This was critical because, an error of on the order of ~1% of a crystal unit cell in the coordinate of the origin (at the impurity atom) produces significant errors in Re$\delta N(\mathbf{q}, E)$ and Im$\delta N(\mathbf{q}, E)$ (supplementary note 1 and Fig. S1). Moreover, single impurity atom based measurements limit the **k**-space resolution because the FOV is typically restricted in size, making them unsuitable for superconductors with large impurity-atom densities. This provides the motivation for a variety of approaches to $\Delta_\mathbf{k}^\alpha$ determination beyond single-atom HAEM. One is to study Bogoliubov bound-states at individual impurity atoms[8,9,10], although this has proven problematic because the elementary HAEM concept (Eq.(3)) is only valid in the weak scattering range i.e. well below the scattering strength sufficient to generate Bogoliubov bound states[11]. Another approach is to use sparse blind deconvolution[12] to analyze images of scattering interference at multiple atoms, yielding the phase-resolved real space structure of $\delta N(\mathbf{r}, E)$ although not the $\rho^-(\mathbf{q}, E)$ of Eq.(3). Overall, therefore, widespread application of the HAEM technique (Eq.(3)) as a general tool for $\Delta_\mathbf{k}^\alpha$ determination remains a challenge.

Here, we introduce a practical technique for determining $\rho^-(\mathbf{q}, E)$ of Eq.(3) from multiple impurity atoms in a large FOV. To understand this approach, consider the key issue of phase analysis as depicted in Fig. 1, a schematic simulation of Friedel oscillations $\delta N(\mathbf{r}) = I_0 \sum_{\mathbf{R}_i} \cos(2\mathbf{k}_F \cdot (\mathbf{r} - \mathbf{R}_i) + \vartheta)/|\mathbf{r}-\mathbf{R}_i|^2$ from multiple atoms at random locations $\mathbf{R}_i$. The Fourier transform components of this $\delta N(\mathbf{r})$ are shown in in the top two panels of Fig. 1b. Obviously, the Re$\delta N(\mathbf{q})$ required for the HAEM technique in Eq.(3), is weak, does not have a clear sign, and is indistinguishable from Im$\delta N(\mathbf{q})$. Such effects occur because the spatial phases of all the individual Friedel oscillations at $\mathbf{R}_i$ are being added at random. The consequence is most obvious in the azimuthally integrated Re$\delta N(\mathbf{q})$ shown in Fig. 1f where the phase information of single-atom Friedel oscillation is completely scrambled and the HAEM technique of Eq.(3) thereby rendered inoperable.

This problem could be mitigated if the Fourier transform of the scattering interference pattern surrounding each $\mathbf{R}_i$ were evaluated as if it were at the zero of coordinates. In this regard consider the Fourier transform of a scattering interference surrounding a single impurity atom at $\mathbf{R}_i = (x_i, y_i)$,

$$\int \delta N(\mathbf{r}-\mathbf{R}_i)e^{i\mathbf{q}\cdot\mathbf{r}}d\mathbf{r} = e^{i\mathbf{q}\cdot\mathbf{R}_i}\int \delta N(\mathbf{r}-\mathbf{R}_i)e^{i\mathbf{q}\cdot(\mathbf{r}-\mathbf{R}_i)}d(\mathbf{r}-\mathbf{R}_i) = e^{i\mathbf{q}\cdot\mathbf{R}_i}\delta N(\mathbf{q}) \qquad (5)$$

This 'shift theorem' shows how the correctly phase-resolved Fourier transform of a $\delta N_i(\mathbf{r})$ oscillation centered on an atom located at $\mathbf{R}_i = (x_i, y_i)$, can be determined using

$$\delta N_i(\mathbf{q}) = e^{i\mathbf{q}\cdot\mathbf{R}_i}\delta N(\mathbf{q}) \qquad (6)$$

where $\delta N(\mathbf{q})$ is the Fourier transform using the same arbitrary origin as determines the $\mathbf{R}_i$. Thus we may define a multi-atom phase-preserving algorithm for QPI

$$\delta N_{\mathrm{MA}}(\mathbf{q}) = \sum_{\mathbf{R}_i} \delta N_i(\mathbf{q}) = \delta N(\mathbf{q}) \sum_{\mathbf{R}_i} e^{i\mathbf{q}\cdot\mathbf{R}_i} \qquad (7)$$

The consequences of Eq.(7) are illustrated in–Fig. 1d and e (supplementary note 3 and supplementary Fig. S4). The real part $\mathrm{Re}\delta N_{\mathrm{MA}}(\mathbf{q})$ now becomes well-defined and the overall magnitude is also strongly enhanced compared to $\mathrm{Re}\delta N(\mathbf{q})$. Moreover, the azimuthally integrated $\mathrm{Re}\delta N_{\mathrm{MA}}(\mathbf{q})$ plotted in Fig. 1g shows that the sign of $\mathrm{Re}\delta N_{\mathrm{MA}}(\mathbf{q})$ changes for $\vartheta = 0$ and $\vartheta = \pi$ as expected. Here it is essential that the impurity atom coordinates $\mathbf{R}_i$ be determined accurately so that the phase is well-defined. We therefore employ a picometer-scale transformation[13,14,15] which renders topographic images $T(\mathbf{r})$ perfectly periodic with the lattice, and then use the same transformation on the simultaneously recorded $g(\mathbf{r}, E)$ to register all the scattering interference oscillations precisely to the crystal lattice (supplementary note 2).

Equation 7 then allows to correctly define the quantities in Eq.(3) for arbitrarily large numbers of scattering atoms. By using the analog of Eq.(6) for $g(\mathbf{r}, E) \propto \delta N(\mathbf{r}, E)$, , $\rho^-(\mathbf{q}, E)$ for each impurity atom is determined from

$$\rho_i^-(\mathbf{q}, E) \propto \mathrm{Re}\{g(\mathbf{q}, +E)e^{i\mathbf{q}\cdot\mathbf{R}_i}\} - \mathrm{Re}\{g(\mathbf{q}, -E)e^{i\mathbf{q}\cdot\mathbf{R}_i}\} \qquad (8)$$

while from Eq.(7) the sum over these $\rho_i^-(\mathbf{q}, E)$ yields

$$\rho_{MA}^-(\mathbf{q}, E) = \sum_i \rho_i^-(\mathbf{q}, E) \qquad (9)$$

This procedure adds all the individual $\rho_i^-(\mathbf{q}, E)$ signals from every impurity atom at $\mathbf{R}_i$ in-phase, while effectively averaging out the random phase variations due to both locating the origin and the contributions of all other scatterers (supplementary Fig. S5). We designate this procedure multi-atom HAEM (MAHAEM).

## RESULTS AND DISCUSSIONS

### Multi-Atom Quasiparticle Interference for $\Delta_\mathbf{k}^\alpha$ Determination

Determination of the magnitude of superconducting energy gaps $|\Delta_\mathbf{k}^\alpha|$ has long been achieved[16-23] using quasiparticle scattering interference (QPI). MAHAEM is the most recent advance of the QPI technique, and to test it we consider FeSe where the single impurity atom HAEM technique for determining $\Delta_\mathbf{k}^\alpha$ was established experimentally[6]. We measure the differential tunneling conductance $g(\mathbf{r}, E) \equiv dI/dV(\mathbf{r}, E)$ in a 30 nm FOV at T=280 mK, followed by determination of $\mathbf{R}_i = (x_i, y_i)$ for 17 scattering sites (supplementary note 3), some of which are shown in the FOV in Fig. 2a (supplementary Fig. S2 shows all the sites.). These sites are well-known Fe-atom vacancies identified by their crystal locations, and are non-magnetic[6]; their empirical identicality is confirmed by high-resolution electronic structure imaging. We then use Eq.(9) to calculate $\rho_{MA}^-(\mathbf{q}, E)$. Figure 2b shows the FeSe Fermi surface with the hole-pocket $\alpha$ around Γ-point and electron pockets $\varepsilon(\delta)$ around X(Y) points. Scattering between $\alpha$ and $\varepsilon$ at wavevector $\mathbf{p}_1$ was studied. A representative layer $\rho_{MA}^-(\mathbf{q}, E = 1.05 \text{ meV})$ is shown in Fig. 2c, where the scattering feature at vector $\mathbf{p}_1$ is marked with a circle. We then sum over the encircled $\mathbf{q}$-region to get $\rho_{MA}^-(E)$ for this scattering feature which is shown as black dots in Fig. 2d. Results from our MAHAEM measurements agree very well with the experimental results using a single impurity atom $\rho_{Single}^{-Exp}$ (black crosses) and the theoretically predicted curve for $\rho_{s_\pm}^{-Th}$ (solid, black) in FeSe. This demonstrates the validity and utility of the multi-atom HAEM technique.

Next we consider LiFeAs, a complex iron-based superconductor that is a focus of contemporary physics interest[24,25,26], particularly the relative sign of $\Delta_\mathbf{k}^\beta$ between all five bands. Fig. 3b shows the Fermi surface of LiFeAs calculated using a tight-binding fit[27,28] to the experimental data. It consists of three hole pockets $h_1$, $h_2$, and $h_3$ around Γ-point and 2 electron pockets $e_1$ and $e_2$ around X-point. The hole pockets around Γ–point on the Fermi surface (FS) revealed by spectroscopic imaging scanning tunneling microscope (SI-STM)[18] and confirmed by angle resolved photoemission spectroscopy (ARPES)[29,30], are much smaller as compared to most other Fe-based superconductors. Local density approximation (LDA) and dynamical mean field theory (DMFT) calculations have attributed the small size of hole pocket to stronger electron-electron correlation in this material. The superconducting energy-gaps $\Delta_\mathbf{k}^\alpha$ are substantially anisotropic[18]. Theoretically, in the case of $\Delta_\mathbf{k}^\alpha$ with $s_\pm$ symmetry, if both electron-like and hole-like pockets are present[31,32] the pairing arises from spin-fluctuations which are enhanced by nesting between the electron-like and hole-like pockets. But the presence of three hole pockets, combined with relatively weak spin fluctuations[33], allow for several possible competing ground states in the presence of repulsive interactions. In Ref. 34 it was pointed out that, under these conditions, several s-wave channels are nearly degenerate. These channels include the $s_\pm$ state where the signs on all hole pockets are the same[35,36] and opposite to the signs on the electron bands, so-called "orbital antiphase state" that occurs when the interaction is diagonal in orbital space[24], and a distinct sign structure obtained when vertex corrections were included[36]. Ref. 37 considered the question of whether these various proposed phases could be distinguished using HAEM based on Eq.(3) and concluded that it would be challenging.

Here we examine the relative signs of $\Delta_\mathbf{k}^\alpha$ in LiFeAs by using MAHAEM. Figure 3a shows the typical cleaved surface of LiFeAs. The scattering sites used in our analysis are Fe-atom vacancies which are non-magnetic (supplementary Fig. S3). The theoretical simulations for LiFeAs were performed from the experimentally fitted tight binding model[27] and anisotropic gap magnitude structure[18,30]. At wavevectors corresponding to electron-hole scattering in **q**-space, a "horn-shaped" feature in $g(\mathbf{q}, E)$ appears within which we focus on an exemplary scattering vector $\mathbf{q}_{\text{eh}}$ indicated by a dashed arrow in Fig. 3b. Figure 3d then shows the theoretical, single-atom $\rho^{-\text{Th}}(\mathbf{q}, E)$ integrated for the **q** in the brown oval in Fig. 3c for $s_\pm$ and $s_{++}$ gaps, where sign of the gap was imposed by hand. The sign of $\rho_{s_\pm}^{-\text{Th}}$ doesn't change for the energy values within the superconducting gap and its amplitude peaks at the energy $E \approx \Delta^{e_1}\Delta^{h_1}$, both characteristics of a sign changing gap[37]; contrariwise $\rho_{s_{++}}^{-\text{Th}}$ changes sign indicative of same sign energy gaps throughout.

For comparison, differential conductance $g(\mathbf{r}, E)$ imaging of LiFeAs is performed at T=1.2K. The typical $g(E)$ spectrum consists of two gaps corresponding to $\Delta_1 = 5.3$ meV and $\Delta_2 = 2.6$ meV. The measured $g(\mathbf{q}, E)$ are shown in Fig. 4a. and the feature at $\mathbf{q}_{\text{eh}}$ expected from the theoretical model in Fig. 3c is indicated by a circle. We evaluate $\rho_{\text{MA}}^-(\mathbf{q}, E)$ from Eq.(9) for N=100 atomic scale Fe-atom vacancy sites (supplementary note 4). The resulting image $\rho_{\text{MA}}^-(\mathbf{q}, E)$ at a representative subgap energy $E = 3.3$ meV is shown in Fig. 4b.

Of note in Fig. 4b is the variety of structures at $|\mathbf{q}| \ll |\mathbf{q}_{\text{eh}}|$, which are challenging to understand. The thin outer blue ring (indicated by dashed light blue curve as guide-to-eye) is located at a radius in $\mathbf{q}$-space that corresponds well to the expected intraband scattering within pocket h3. Furthermore, much of the $\mathbf{q}$-space within this ring is blue and of rather high intensity for 1 meV<|E|<6 meV (supplementary Fig. S6a shows dashed contours for various possible inter-hole-band scatterings overlaid on the unprocessed $\rho^-(\mathbf{q}, E)$). The blue color, indicating sign-preserving scattering, is consistent with the conventional $s_\pm$ picture within a HAEM scenario, but the high intensity is not. As discussed in supplementary note 5 there are several possible explanations of these low $|\mathbf{q}|$ phenomena, including strong scattering, quasiparticle bound states and antiphase hole-pocket gaps.

Nevertheless, when the high $|\mathbf{q}|$ scattering between hole-like and electron-like pockets (Fig. 3b and Fig. 3c) is integrated within the $\mathbf{q}$-space region shown by a brown circle on the $\rho_{\text{MA}}^-(\mathbf{q}, E)$ of Fig. 4a, it yields $\rho_{\text{MA}}^-(E)$ as plotted in Fig. 4c. The theoretically predicted $\rho^-(E)$ curves are overlaid for comparison. It is clear that the experimental $\rho_{\text{MA}}^-(E)$ is consistent with the $\rho_\pm^{-\text{Th}}(E)$ theory because it does not change sign and exhibits a peak at $E \approx 3.7$ meV $\approx \sqrt{\Delta_1 \Delta_2}$. In this way, the MAHAEM technique efficiently demonstrates that $\Delta_\mathbf{k}^\alpha$ changes sign between electron-like and hole-like bands of LiFeAs.

**Conclusions**

We report development and demonstration of an improved approach for signed $\Delta_{\mathbf{k}}^{\alpha}$ determination (Eq.(9)), but now for use with multiple impurity atoms or scattering centers. This MAHAEM technique for measuring $\rho^{-}(\mathbf{q}, E)$ is based on a combination of the Fourier shift theorem and high precision registry of scatterer locations. It extends the original HAEM approach[4] to more disordered superconductors (Figs 2a, 3a), enables its application to far larger fields of view thereby enhancing **q**-space resolution (Fig. 4b), and greatly increases signal to noise ratios (Figs 1d, 4b) by suppressing phase randomization in multi-atom scattering interference. Overall, MAHAEM now represents a powerful and general technique for $\Delta_{\mathbf{k}}^{\alpha}$ determination in complex superconductors.

# METHODS

## Sample Growth and Preparation

FeSe samples with $T_c \approx 8.7K$ were prepared using KCl$_3$/AlCl$_3$ chemical-vapour transport and LiFeAs samples with $T_c \approx 15$ K were grown using LiAs flux method. The highly reactive LiFeAs samples are prepared in a dry nitrogen atmosphere in a glove box.

## SI-STM Measurements and Analysis

All samples are cleaved in-situ in our ultra-high cryogenic vacuum STM at low temperature. The $g(\mathbf{r}, E)$ data was acquired with a ³He refrigerator equipped STM. The picometer level atomic registration was performed before applying the HAEM technique as described in full detail in the supplementary note 2. Full details of the multi-atom HAEM analysis are presented in detail in supplementary note 3. Theoretical predictions for $\rho^-(E)$ curves were performed using the T-matrix formalism with energy gap on each band and normal state tight binding parameters fitted to experiments.

**Data Availability:** The datasets generated and/or analysed during this study are available to qualified requestors from the corresponding author.

**Code Availability:** The simulation code for Fig. 1 is provided as supplemental material. All the other codes used during the current study are available to qualified requestors from the corresponding author.

**Acknowledgements:** Work done by P.C.C. and A.E.B. was supported by the U.S. Department of Energy, Office of Basic Energy Science, Division of Materials Sciences and Engineering and was performed at the Ames Laboratory. Ames Laboratory is operated for the U.S. Department of Energy by Iowa State University under Contract No. DE-AC02-07CH11358. RS acknowledges support from Cornell Center for Materials Research with funding from the NSF MRSEC program (DMR-1719875). The authors are thankful to M.A. Müller for the discussion of the QPI results in LiFeAs. P.J.H. and M.A.S. acknowledge support from NSF-DMR-1849751; H.E. acknowledges Grant-in-Aid for Scientific Research on Innovative Areas "Quantum Liquid Crystals" (KAKENHI Grant No. JP19H05823) from JSPS of Japan. J.C.S.D. acknowledges support from the Moore Foundation's EPiQS Initiative through Grant GBMF9457, from the Royal Society through Award R64897, from Science Foundation Ireland

**Author Contributions:** R.S., A.Kr. and M.A.S. contributed to this project equally. P.O.S., R.S., P.J.H and J.C.S.D. designed the project. P.O.S. and M.A.S. developed the phase-resolved multi atom averaging method; M.P.A., A.Ko. and P.O.S. carried out the experiments; R.S. and P.O.S carried out the data analysis; A.Kr., M.A.S., J.B., P.J.H and I.E. carried out the theoretical analysis. P.C.C. and A.E.B. synthesized single crystalline FeSe samples; H.E. synthesized single crystalline LiFeAs samples. J.C.S.D. and P.J.H supervised the investigation and wrote the paper with key contributions from P.O.S., R.S., M.A.S. and A.Kr. The manuscript reflects the contributions of all authors.

**Competing Interests:** The authors declare no competing financial or non-financial interests.

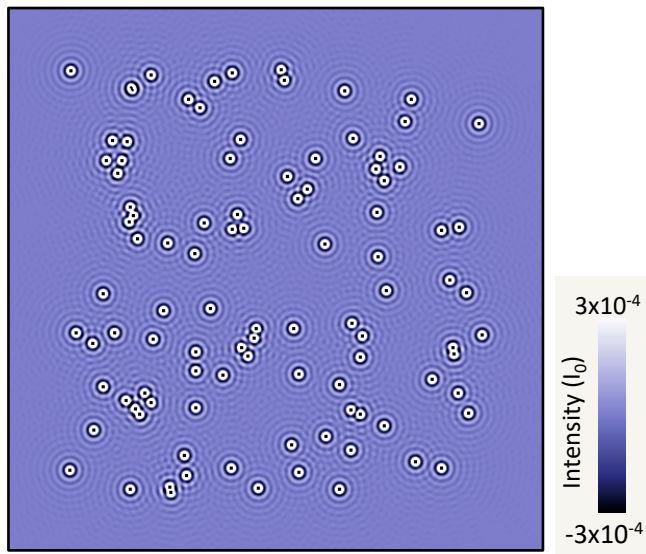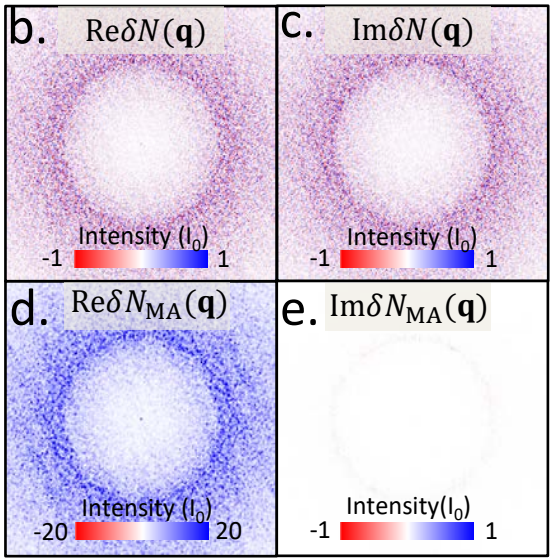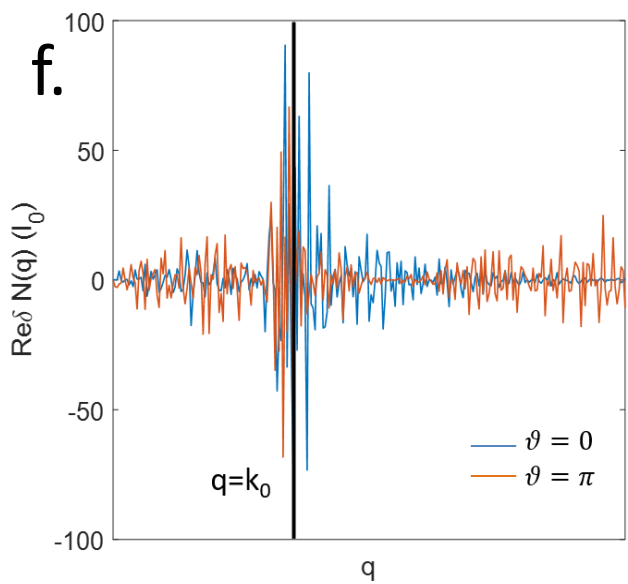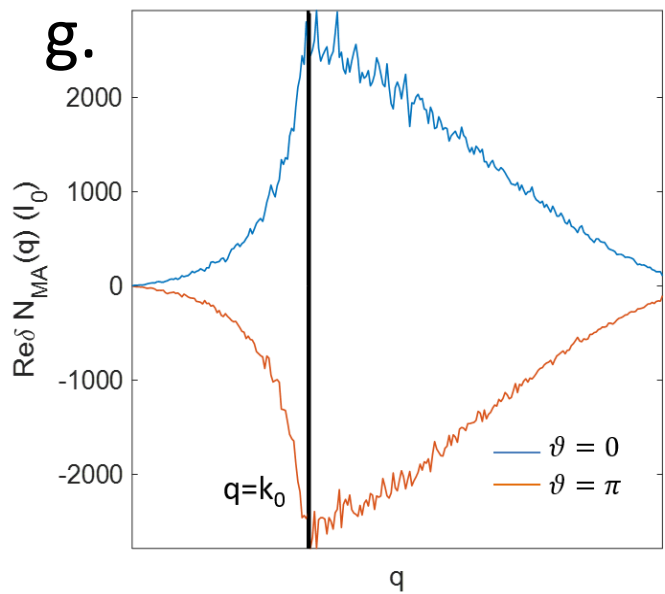

# Figure Captions

1. **Schematic for Multi-Atom Phase Analysis**
   a. Simulation of density of states perturbation $\delta N(\mathbf{r})$ in amplitude units $I_0$ due to 2-dimensional Friedel oscillations surrounding 100 impurity atoms at random locations $\mathbf{R}_i$.
   b. Real part of Fourier transform $\text{Re}\delta N(\mathbf{q})$ from $\delta N(\mathbf{r})$ in 1a. We use an integer grid, hence the units of Fourier transform are also $I_0$.
   c. Imaginary part of Fourier transform $\text{Re}\delta N(\mathbf{q})$ from $\delta N(\mathbf{r})$ in 1a.
   d. Real part of Fourier transform $\text{Re}\delta N_{\text{MA}}(\mathbf{q})$ calculated using multi-atom technique of Eq.(7).
   e. Imaginary part of Fourier transform $\text{Re}\delta N_{\text{MA}}(\mathbf{q})$ calculated using multi-atom technique of Eq.(7).
   f. $\text{Re}\delta N(\mathbf{q})$ from $\delta N(\mathbf{r})$ in 1a for $\vartheta = 0$ and $\vartheta = \pi$, integrated azimuthally from 1b. Its strong random fluctuations versus $|\mathbf{q}|$ are due to summing the Friedel oscillations in $\delta N(\mathbf{r})$ of 1a with random phases due to the random locations $\mathbf{R}_i$.
   g. $\text{Re}\delta N_{\text{MA}}(\mathbf{q})$ from $\delta N(\mathbf{r})$ in 1a integrated azimuthally from-1d. $\text{Re}\delta N_{\text{MA}}(\mathbf{q})$ is now orders of magnitude more intense than in 1f, and the phase of the Friedel oscillations in $\delta N(\mathbf{r})$ of 1a is now very well defined because the effects of random locations $\mathbf{R}_i$ are removed by using Eq.(7). Note that, now, changing the oscillation phase $\vartheta = 0$ to and $\vartheta = \pi$ surrounding all $\mathbf{R}_i$ in $\delta N(\mathbf{r})$ produces the correct evolution of $\text{Re}\delta N_{\text{MA}}(\mathbf{q})$.

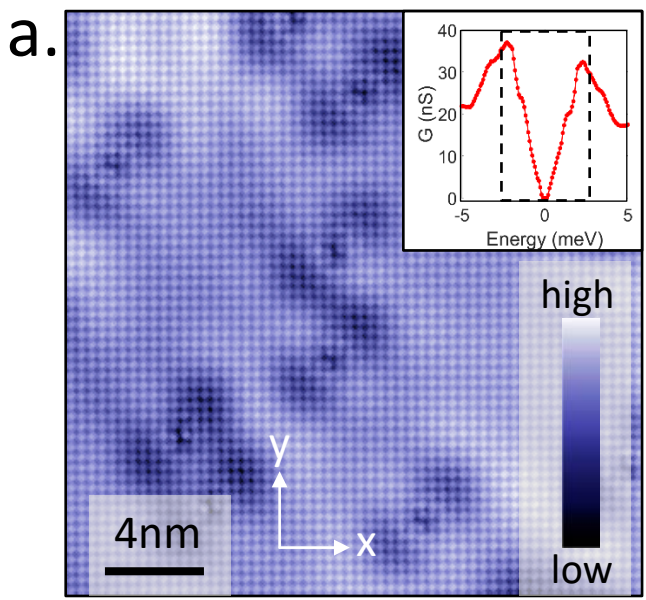
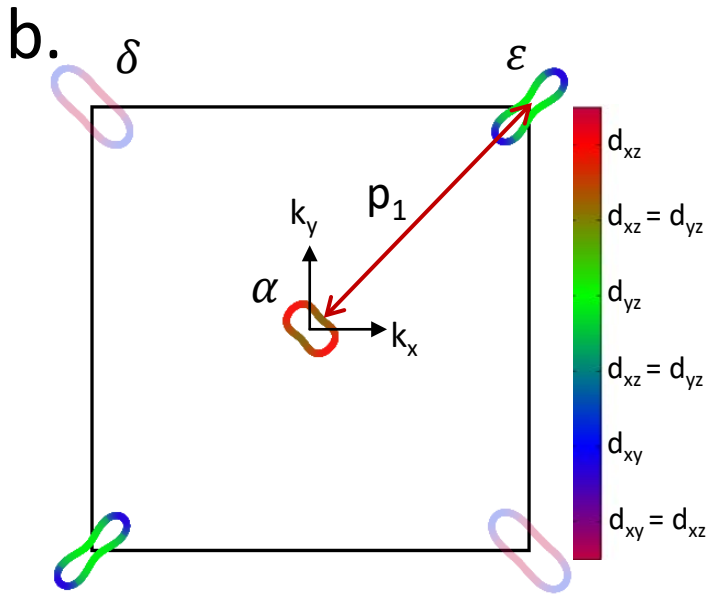
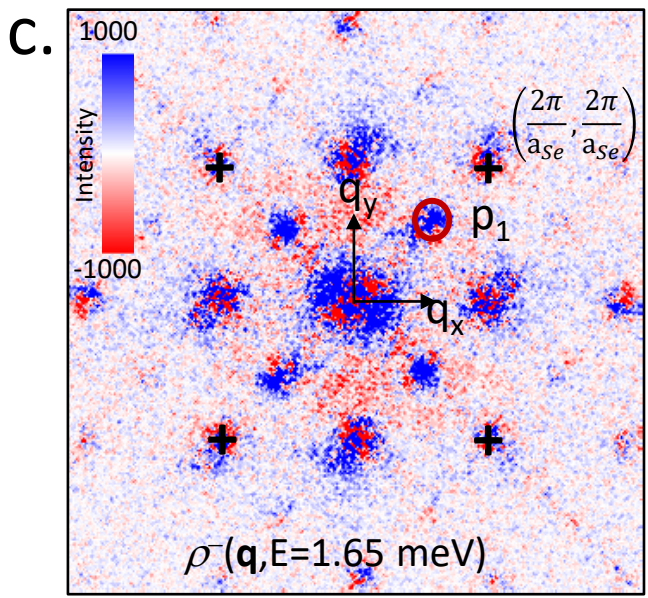
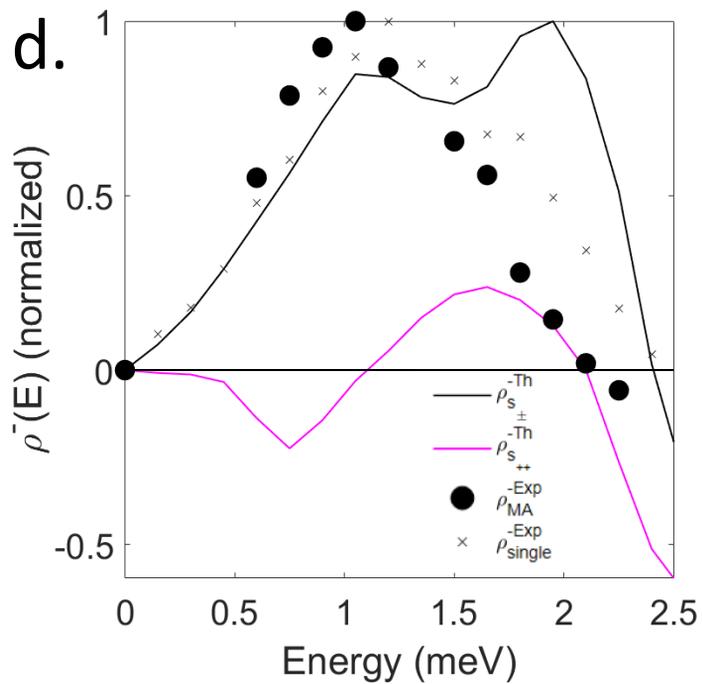

## 2. Demonstration of MAHAEM for FeSe

a. Topography of FeSe showing the type of defects (Fe-atom vacancies) used for analysis. The x- and y-axes are along top Se-Se atoms. Inset shows the differential conductance spectrum recorded at a point on superconducting FeSe. The dashed rectangle represents the energy limits of the high-resolution dI/dV maps used herein for $\rho^-(\mathbf{q}, E)$ analysis.

b. Fermi surface of FeSe showing the scattering between hole-pocket $\alpha$ and electron-pocket $\varepsilon$ with scattering vector $\mathbf{p_1}$, which is the subject of study. The delta pocket is predicted in LDA calculations hence it is shown dim in the image, but is now reported not to exist in reality[6,38]. Due to the orbital content, the scattering between quasi-parallel Fermi surfaces would be strongly suppressed in this orbitally selective material[6].

c. $\rho^-_{MA}(\mathbf{q}, E = 1.05 \text{ meV})$ calculated using Eq.(9) for a FOV containing 17 Fe-vacancies. The circle denotes the region where the $\alpha \rightarrow \varepsilon$ scattering occurs and we integrate the $\rho^-_{MA}(\mathbf{q}, E)$ over the $\mathbf{q}$ in this region. Black crosses denote the Bragg peaks.

d. The integrated $\rho^-_{MA}(E)$ (dots, black) from our MAHAEM analysis of FeSe compare to the theoretical predictions from an accurate band- and gap-structure model of FeSe for $s_{++}$ (solid, pink) and $s_{\pm}$ (solid, black) superconducting energy gap symmetry, and to measured $\rho^{-\text{Exp}}_{\text{single}}(E)$ (crosses, black) from single impurity analysis as reported in Ref. 6. Clearly, the single atom $\rho^{-\text{Exp}}_{\text{single}}(E)$ and the MAHAEM $\rho^-_{MA}(E)$ are in good agreement. Note that the 2D plots may show both red and blue colors due to non-ideal nature of real experimental data. However, subsequent to the integration over relevant $\mathbf{q}$-space region, the $\rho^-_{MA}(E)$ is well defined as demonstrated here.

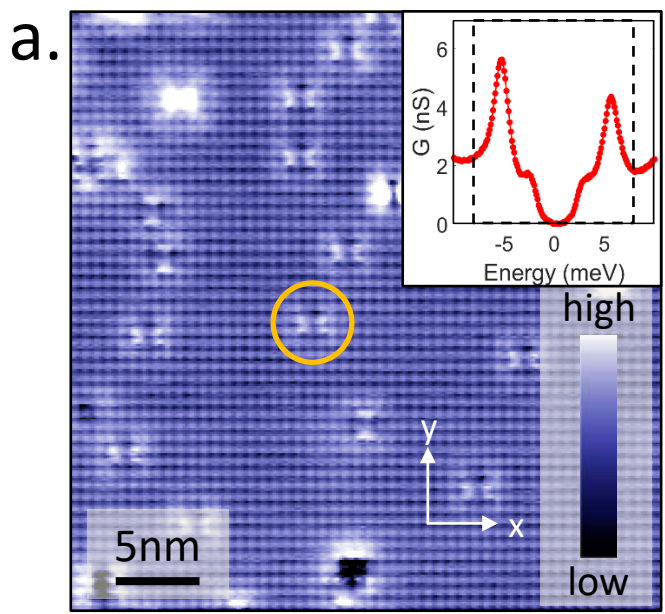
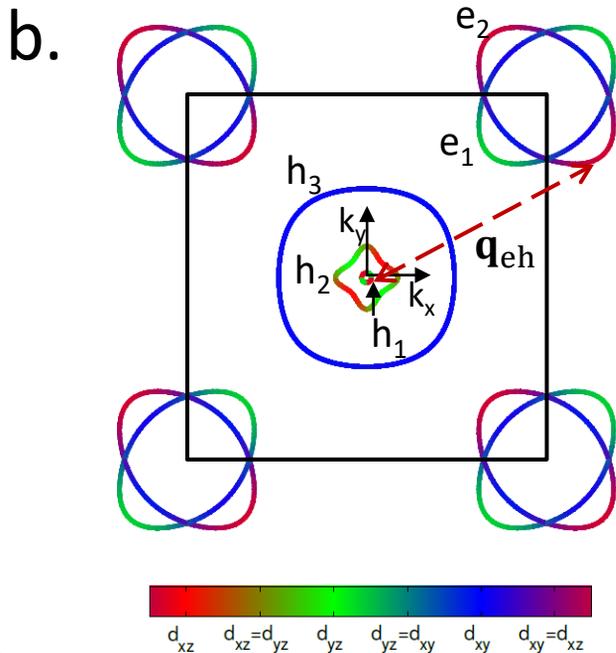
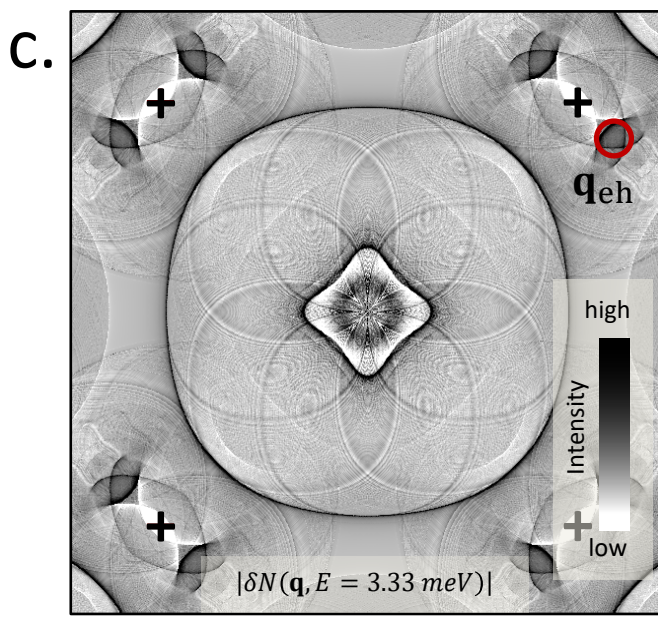
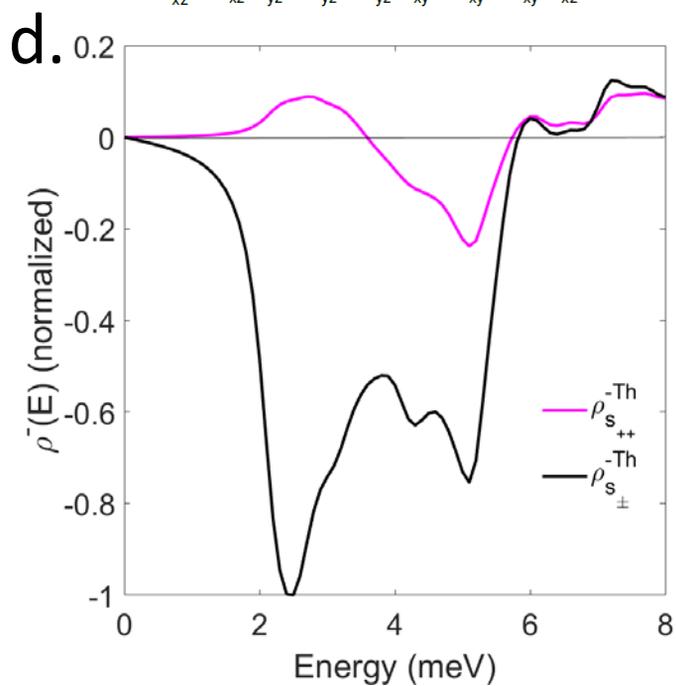

### 3. LiFeAs Scattering Interference

e. Topograph recorded at a LiFeAs surface showing Fe-atom vacancies. The x-axis and the y-axis directions are along As-As directions. The orange circle denotes the type of weak scatterer we chose for our MAHAEM analysis. Inset shows the $g(E)$ spectrum measured at a point for superconducting LiFeAs. The dashed rectangle represents the energy limits of the high-resolution dI/dV maps used herein for $\rho^-(\mathbf{q}, E)$ analysis.

b. The Fermi surface model for LiFeAs showing three hole pockets $h_1$, $h_2$ and $h_3$ around Γ-point and two electron pockets $e_1$ and $e_2$ around X-point in a 2-Fe zone. The scattering from hole-like to electron like pockets takes place as indicated by a dashed vector $\mathbf{q}_{eh}$.

c. Theoretical prediction for single atom $|\delta N(\mathbf{q}, E = 3.25 \text{ meV})|$ using band- and gap-structure values fitted from experiments[18]. The electron-hole scattering near $\boldsymbol{q}_{eh}$ appears as a "horn"-shaped feature which is enclosed by a circle.

d. Theoretical prediction for single atom $\rho^-(E)$ integrated over the circular region shown in Fig. 3c for both $s_\pm$ (black) and $s_{++}$ (pink) symmetry.

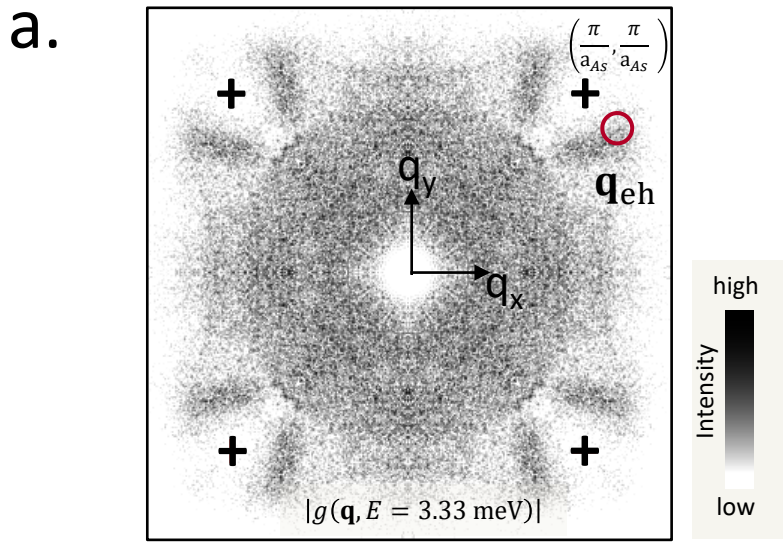

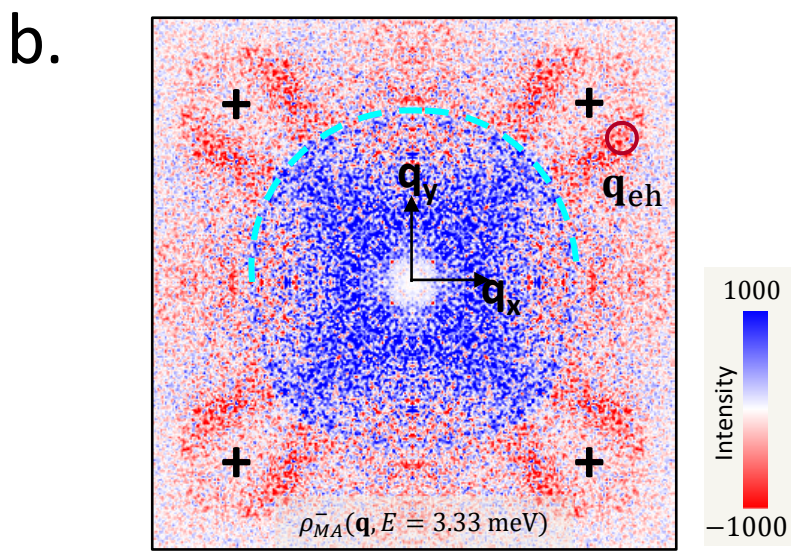

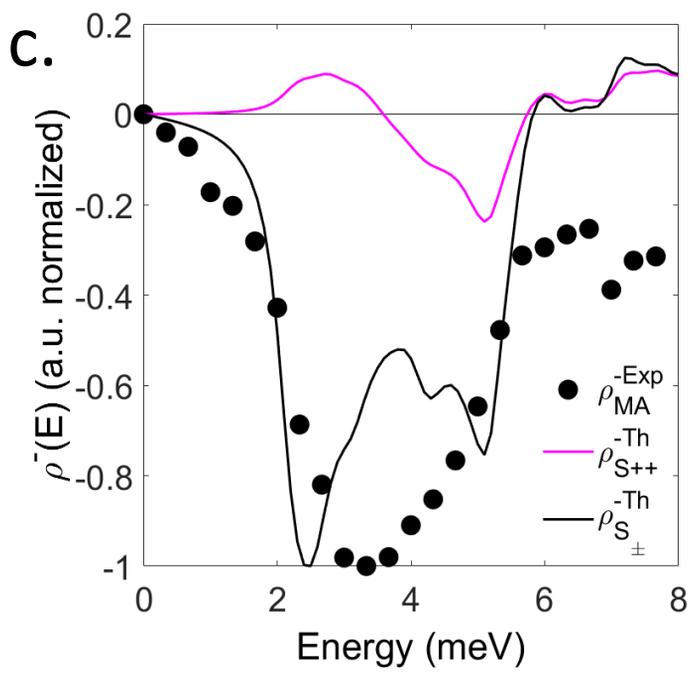

## 4. LiFeAs Superconducting Energy-gap Symmetry from MAHAEM

a. The measured $|g(\mathbf{q})|$ pattern recorded in the FOV with multiple atomic scattering sites. The hole-like to electron-like scattering as predicted in Fig. 3c is detected clearly and indicated by a brown circle at the same location as 3c. The features inside the circle do not appear identical because the intensity in real experimental data falls off at higher $|\mathbf{q}|$, while the theoretical simulation is replicated about the q-space Brillouin zone boundary with the same intensity, creating a non-physical equal intensity reflected feature. A Gaussian mask of $\sigma = 0.68\ \text{Å}^{-1} = 0.8\pi/a$ is used to suppress the $|g(\mathbf{q})|$ data in the $|\mathbf{q}|\approx 0$ core region, to allow clearer presentation of the high-$|\mathbf{q}|$- under study. Much of this $|\mathbf{q}|\approx 0$ signal intensity is believed to emanate from long range disorder and is, moreover, unrelated to the scientific objectives of this paper.

b. The measured $\rho^-_{MA}(\mathbf{q}, E = 3.33\text{ meV})$ using Eq.(9); it is typical of all $\rho^-_{MA}(\mathbf{q}, E)$ between 1 and 6 meV. The circle indicates the hole-like to electron-like scattering in 4a. We integrate the $\rho^-_{MA}(\mathbf{q}, E)$ over the range of $\mathbf{q}$ within this region. The dashed lines are guide to eye to a feature which is consistent with intra $h_3$ scattering. Again, a Gaussian mask of $\sigma = 0.68\ \text{Å}^{-1} = 0.8\pi/a$ is used to suppress the intense core emanating from long range disorder, to allow clearer presentation of the $\rho^-_{MA}(\mathbf{q}, E)$ information at high-$|\mathbf{q}|$. This suppressed area corresponds to $\sim$18% of the total area of the first $\mathbf{q}$-space Brillouin zone, and the unprocessed $\rho^-_{MA}(\mathbf{q}, E)$ data for this figure is provided in its entirety at supplementary Fig. S6a.

c. The resulting $\rho^-_{MA}(E)$ calculated by summing over the oval enclosed region in 4b (black dots), and the theory curves (solid) for $s_\pm$ (black) and $s_{++}$ (pink) overlaid. Clearly, this demonstrates using MAHAEM that the superconducting energy gap symmetry of LiFeAs is $s_\pm$ (black).

Supplementary information for

# Multi-atom quasiparticle scattering interference
# for superconductor energy-gap symmetry determination

## Supplementary note 1

In this note, we describe **phase effects of displacement of impurity atom from origin**. For an atomic impurity located at the origin of the FOV, the scattering interference amplitude $\delta N(\mathbf{r})$ is an even function under inversion with respect to the origin of the FOV. Thus, its Fourier transform $\delta N(\mathbf{q})$ would be completely real (a well-known theorem of Fourier transform). This is demonstrated Fig. S1. In top half of Fig. S1a, scattering interference intensity from an impurity located at the origin of the FOV is simulated as a 2D Friedel oscillation $\delta N(\mathbf{r}) = I_0 \frac{\cos(2\mathbf{k}_F \cdot \mathbf{r} + \vartheta)}{|\mathbf{r}|^2}$ on a NxN integer grid representing $\mathbf{r}$. The unprocessed Fourier transform shown in Fig. S1b and c shows that the signal is in the real channel. The azimuthally integrated real part of the Fourier transform Re $\delta N(\mathbf{q})$ is plotted in Fig. S1f and shows a clear sign which is opposite for $\vartheta = 0$ and $\vartheta = \pi$ as expected and these cases can be distinguished easily.

However, if the impurity center is shifted by $\mathbf{R}_0$ with respect to the origin of the FOV, then the scattering interference signal is no longer an even function with respect to the origin. By the shift theorem of Fourier transform, the Fourier transform of the image with the impurity shifted from the origin of the FOV is given as:

$$\delta N_S(\mathbf{q}) = e^{i\mathbf{q}\cdot\mathbf{R}_0} \delta N_S(\mathbf{q}) \tag{1}$$

$$\text{Re}\delta N_S(\mathbf{q}) = \cos(\mathbf{q}\cdot\mathbf{R}_0)\,\delta N(\mathbf{q}), \quad \text{Im}\delta N_S(\mathbf{q}) = \sin(\mathbf{q}\cdot\mathbf{R}_0)\delta N(\mathbf{q}) \tag{2}$$

Hence, the Fourier transform of the scattering interference signal from an impurity shifted with respect to origin of the FOV now contains both real and imaginary part which oscillate with frequency $\mathbf{R}_0$ as a function of $\mathbf{q}$. This can be seen in Fig. S1. In bottom half of Fig. S1a, scattering interference intensity from an impurity shifted by $\mathbf{R}_0 = (x_0, y_0)$ ($x_0$ and $y_0$ are random integers picked on a NxN grid) from the origin of the FOV is simulated as a Friedel oscillation $\delta N(\mathbf{r}) = I_0 \frac{\cos(2\mathbf{k}_F \cdot (\mathbf{r} - \mathbf{R}_0) + \vartheta)}{|r - \mathbf{R}_0|^2}$. The unprocessed Fourier transform shown in Fig. S1d

and e now shows that the signal is now oscillating rapidly between positive and negative (red and blue) and the magnitude is distributed between both real and imaginary parts. The azimuthally integrated real part of the Fourier transform Re $\delta N_S(\mathbf{q})$ is plotted in Fig. S1g and this reflects the rapid oscillations with no clear sign and therefore no distinction can be made in the Re $\delta N_S(\mathbf{q})$ from $\vartheta = 0$ and Re $\delta N_S(\mathbf{q})$ from $\vartheta = \pi$.

## Supplementary note 2

In this note, we describe how an **atomic scale registration to a perfect crystal lattice** is achieved. An atomic scale perfect topograph with orthogonal unit cell vectors **a** and **b** recorded by STM can be represented as

$$T(\mathbf{r}) = T_0[\cos(\mathbf{Q}_a \cdot \mathbf{r}) + \cos(\mathbf{Q}_b \cdot \mathbf{r})] \qquad (3)$$

where $\mathbf{Q}_a = (Q_{ax}, Q_{ay})$ and $\mathbf{Q}_b = (Q_{bx}, Q_{by})$ are the two Bragg wavevectors at which the atomic modulations occur. In a real experiment recorded, $T(\mathbf{r})$ may suffer from a slowly varying distortion[1,2,3] $u(\mathbf{r})$. Hence the topographic image in a real experiment can be written as:

$$\tilde{T}(\mathbf{r}) = T_0[\cos(\mathbf{Q}_a \cdot \tilde{\mathbf{r}}) + \cos(\mathbf{Q}_b \cdot \tilde{\mathbf{r}})] \qquad (4)$$

Where $\tilde{\mathbf{r}} = \mathbf{r} + \mathbf{u}(\mathbf{r})$. This distortion leads to an additional phase at each location **r** given by

$$\begin{pmatrix} \theta_a(\mathbf{r}) \\ \theta_b(\mathbf{r}) \end{pmatrix} = \mathbf{Q}\mathbf{u}(\mathbf{r}) \qquad (5)$$

Where $\mathbf{Q} = (\mathbf{Q}_a^T, \mathbf{Q}_b^T)$ is an orthogonal matrix which is invertible, allowing one to solve for the displacement field $\mathbf{u}(\mathbf{r})$ as:

$$\mathbf{u}(\mathbf{r}) = \mathbf{Q}^{-1} \begin{pmatrix} \theta_a(\mathbf{r}) \\ \theta_b(\mathbf{r}) \end{pmatrix} \qquad (6)$$

To find $\theta_i$, we employ a computational lock-in technique in which the topograph, $T(\mathbf{r})$, is multiplied by reference sine and cosine functions with periodicity set by $\mathbf{Q}_a$ and $\mathbf{Q}_b$. The resulting four images are filtered to retain only the **q**-space regions within a radius $\delta q = \frac{1}{\lambda}$ of the four Bragg peaks; the magnitude of $\lambda$ is chosen to capture only the relevant image distortions. This procedure results in retaining the local phase information $\theta_a(\mathbf{r}), \theta_b(\mathbf{r})$ that quantifies the local displacements from perfect periodicity:

$$Y_i(\mathbf{r}) = \sin\theta_i(\mathbf{r}), \quad X_i(\mathbf{r}) = \cos\theta_i(\mathbf{r}) \tag{7}$$

Dividing the appropriate pair of images allows one to extract $\theta_i(\mathbf{r})$:

$$\theta_i(\mathbf{r}) = \tan^{-1}\frac{Y_i(\mathbf{r})}{X_i(\mathbf{r})} \tag{8}$$

Once we have $\theta_i(\mathbf{r})$, we get the displacement field $\mathbf{u}(\mathbf{r})$ using Eq.(6) and the value of recorded topograph $\tilde{T}(\mathbf{r})$ as given by Eq.(4) can now be registered to a perfect ideal lattice $T(\mathbf{r})$ as given by Eq.(3) using:

$$T(\mathbf{r}) = \tilde{T}(\mathbf{r} - \mathbf{u}(\mathbf{r})) \tag{9}$$

## Supplementary note 3

In this note, we describe **MAHAEM analysis procedure** step-by-step. To get optimal phase resolution, it is essential to determine the scattering center of the defect in the conductance map as precisely as possible. Thus, if the required $\mathbf{q}$-space resolution allows to resolve atoms, we use the technique as mentioned in supplementary note 2 to register each atom in the recorded topograph $T(\mathbf{r})$ map to a perfectly periodic lattice using an affine transformation at a picometer level precision[1,2,3]. Then we apply the same transformation to the recorded conductance maps $g(\mathbf{r}, E)$ for each energy $E$ to register each scattering interference pattern perfectly to the crystal lattice as seen in the topograph.

More than one type of scatterer can be observed in a given field of view. The basic HAEM prescription works correctly only for weak impurities[7] which are therefore our focus. Strong scatterers appear very bright on the topograph and conductance maps in our chosen colorscale and can be ruled out visually. The weak scatterers have very similar visual appearance in topograph e.g. red circle in Fig. 3a. They are also numerically predominant, hence easily chosen visually for the MAHAEM analysis.

From the perfectly registered $g(\mathbf{r}, E)$ images, one records the coordinates of the centers of the impurities. An example for FeSe is shown in Fig. S2a. For example, in FeSe, the Fe-vacancy site which acts as the impurity potential is chosen as the pixel in the middle of the two bright spots which are due to Se atoms. The coordinates chosen by us for analysis are shown by red crosses on Fig. S2a. Using these impurity coordinates $\mathbf{R}_i$ we calculate

Fourier transform as if the impurity is shifted to the origin by using shift theorem of Fourier transform.

$$g_i^S(\mathbf{q}, E) = e^{i\mathbf{q}\cdot\mathbf{R}_i} g(\mathbf{q}, E) \tag{10}$$

The shift theorem of Fourier transform shifts the map with periodic boundary conditions. this can be seen in Fig. S2b, where the impurity in top half of image has been shifted to the center in the bottom half of Fig. S2b. This process may lead to discontinuous edges in the image as seen in the bottom half of Fig. S2b, which can lead to spurious noise features in the Fourier transform. To ameliorate this effect, one could apply a sinusoidal window before performing the shift so that the values go smoothly to zero on the edges and then shifting and adding would not lead to discontinuities. However, this process would weigh the scattering interference near the center more than the scattering interference near the edges.

In the cases of LiFeAs and FeSe, we realized that empirically such edge discontinuities lead to almost negligible effects as compared to the intensity of scattering interference. Hence, for all the analysis in this paper, we inverse Fourier transform $g_i^S(\mathbf{q}, E)$ to get $g_i^S(\mathbf{r}, E)$ and apply a sinusoidal window to the shifted image to get windowed and shifted maps $g_i^{SW}(\mathbf{r}, E)$ as:

$$g_i^{SW}(\mathbf{r}, E) = W g_i^S(\mathbf{r}, E) \tag{11}$$

where $W = \left[\sin\left(\frac{\pi r_x}{L}\right)\right]^T \sin\left(\frac{\pi r_y}{L}\right)$ is the windowing function with L denoting the size of FOV and $\mathbf{r}_x, \mathbf{r}_y$ are vectors with $0 < r_{x,y} < L$. The Fourier transform of $g_i^{SW}(\mathbf{r}, E)$ then yields the functional Fourier transform of each shifted impurity as $g_i(\mathbf{q}, E)$. Mathematically, if $F$ denotes Fourier transform, then this sequence of steps to obtain $g_i(\mathbf{q}, E)$ from recorded $g(\mathbf{r}, E)$ can be represented in a single equation as:

$$g_i(\mathbf{q}, E) = F\left(W F^{-1}\left(e^{i\mathbf{q}\cdot\mathbf{R}_i} g(\mathbf{q}, E)\right)\right) \tag{12}$$

From these $g_i(\mathbf{q}, E)$, we calculate the key HAEM quantity $\rho_i^-(\mathbf{q}, E)$ for impurity $\mathbf{R}_i$ as

$$\rho_i^-(\mathbf{q}, E) = \text{Re}\{g_i(\mathbf{q}, E)\} - \text{Re}\{g_i(\mathbf{q}, -E)\} \tag{13}$$

Finally, the sum of all $\rho_i^-(\mathbf{q}, E)$ from all impurities $\mathbf{R}_i$ yields

$$\rho_{\text{MA}}^-(\mathbf{q}, E) = \sum_{i=1}^{N} \rho_i^-(\mathbf{q}, E) \qquad (14)$$

Hence all noise from out of phase signal from impurities which are not at the origin of the FOV is severely suppressed. Although not used for superconducting gap symmetry determination, similar Fourier analysis concepts were considered in Ref. 4. Whereas the signal from each impurity which has been brought in phase by bringing each impurity to the center using shift theorem, gets added and hence enhanced by a factor of number of impurities $N$. Fig. S3 shows all these steps for LiFeAs in an analogous manner to Fig. S2 for FeSe.

Fig. S4 presents the effectiveness of this procedure. Fig S4a shows an image of $\delta N(\mathbf{r}) = I_0 \sum_{\mathbf{R}_i} \cos(2\mathbf{k}_F \cdot (\mathbf{r} - \mathbf{R}_i) + \vartheta)/|\mathbf{r}-\mathbf{R}_i|^2$ for 20 random sites $\mathbf{R}_i$. We do not consider the interference between impurities at different $\mathbf{R}_i$ because experimentally, the interference pattern disappears at shorter distances compared to the impurity spacing. Fig. S4b shows the real-space sum of all impurities calculated after shifting each impurity to the center with periodic boundary condition. It can be noticed clearly that the intensity for the impurity at the center is enhanced and more rings of Friedel oscillations $\delta N(\mathbf{r})$ can be seen whereas the impurities which are away from the center appear with diminished intensity. The real part of the Fourier transform of Fig. S4a with 20 random impurity centers is oscillatory with no clear sign as shown in Fig. S4c and the intensity is divided between real and imaginary part (Fig. S4d) due to no well-defined origin for phase. Whereas, after applying the MAHAEM procedure and obtaining Fig. S4b, its Fourier transform has a definite sign in the real part as seen in Fig. S4e and imaginary part is much smaller in intensity as seen in Fig. S4f. This demonstrates the utility of this procedure to recover phase for multiple-impurity scenarios.

As a practical example, Fig. S5 shows how our MAHAEM technique recovers $\rho^-(\mathbf{q}, E)$ with multiple real scatterers. Fig. S5a shows the real-space $\sum_{i=1}^{N} g^S(\mathbf{r}, E = 5.33 \text{ meV})$ calculated using Eq.(10) for N=1,10,50 and 100. The central impurity becomes brighter and the impurities elsewhere are averaged out, exactly as the simulations in Fig. S4 predict. Fig. S5b shows symmetrized $\rho_{\text{MA}}^-(\mathbf{q}, E) = \sum_{i=1}^{N} \rho_i^-(\mathbf{q}, E)$ for $E = 5.33$ meV calculated using Eq.(14) for N=1,10,50 and 100 impurities. Here we see that the $\rho_{\text{MA}}^-(\mathbf{q}, E)$ becomes more and more well defined in terms of scattering features having a clear sign (blue=positive, red=negative) as the N increases. Finally, we show the $\rho_{\text{MA}}^-(E)$, integrated over the circle around the vector $\mathbf{q}_{\text{eh}}$ as depicted in Fig. 3b, demonstrating how the signal improves as N increases in the case of real scatterers.

In real experiments for $\rho^-(\mathbf{q}, E)$, the effectiveness of MAHAEM procedure can be seen by comparing Fig. S2d, which shows $\rho^-(\mathbf{q}, E)$ with Fig. S2c, which shows $\rho_i^-(\mathbf{q}, E)$ calculated using a single scatterer shown in Fig. S2b. The circle indicates the electron-hole scattering vector $p_1$ which was studied in Ref. 5 and shown in Fig. 2b of the main text. Black crosses denote Bragg peaks. Fig. S3 shows the MAHAEM enabled phase resolution in an analogous manner. A theoretical calculation using the multi-impurity Green's function method provides the same conclusions as the single-impurity T-matrix method, provided the MAHAEM method is employed[6].

# Supplementary note 4

In this note, we establish the **robustness of MAHAEM in LiFeAs**. The results in the main-text Fig. 4b. are 8-fold symmetrized along vertical, horizontal and diagonal axes for enhanced clarity. However, the symmetry is present in the unprocessed data itself as can be seen in the unprocessed data presented in Fig. S6 for a representative layer $E = 3.33$ meV. The scattering vector from small hole bands around Γ- point to the electron bands at X point leads to the horn-shaped feature as discussed in the main-text. In Fig. S6, we show the $\rho_{\text{MA}}^-(E)$ integrated in the circle of same radius as in Fig. 4c of the main text at different positions on the horn. The $\rho_{\text{MA}}^-(E)$ for all these positions on the horn in the unprocessed data is very similar to the $\rho_{\text{MA}}^-(E)$ presented in the main text in Fig. 4c. Most importantly, it does not change sign and the negative-value peaks at $E \approx \Delta_1, \Delta_2$ as in Fig. 4c. and as expected from detailed theoretical calculations of $\rho^-(E)$ using the experimental Fermi surface and superconducting gaps as outlined in the main-text. .

The distinct two-hump structure from the simulation is not observed in the data as can be noticed in 4c. This can be explained by following two considerations. Firstly, the electronic structure for real systems is not arbitrarily sharp i.e. the spectral functions exhibit a finite broadening which is set to practically zero in the theoretical calculations. Also, tunneling effects are not included in our calculations and thus the contribution due to the $d_{xy}$ electrons, whose contribution to the $\rho^-(E)$ at the inner gap energy (inset Fig, 3a) will qualitatively reduce the peak at smaller energies.

## Supplementary note 5

In this note, we discuss the possibilities for **low |q| MAHAEM in LiFeAs**. As shown in the main text Fig. 4b and also in the unprocessed data in Fig. S6, there is a large intensity in $\rho^-(\omega)$ at small q which additionally has a different overall sign than the MAHAEM signal used to determine the relative sign between the gap on the electron- and hole-like pockets. From a theoretical point of view, there are a number of possible explanations to be considered for future research:

a) *Intermediate-strength scatterers.* As already discussed in the original HAEM paper, the clear distinction between gap-sign-preserving and sign-changing processes is preserved for intermediate strength scatterers, but the intensity of the sign-preserving processes is distinctly enhanced. This, together with a possible form factor in q-space due to the longer-range nature of impurity potentials, could lead to enhanced intensities at small q relative to large q. The intensity might be quite diffuse in q-space, particularly since the inner xz/yz hole pockets are three dimensional, or nearly so.

b) *Impurity bound states.* If impurity potentials are strong enough, bound states will be formed in the gap of a gap-sign-changing system. We do not know the identity of the native defects in our LiFeAs samples, but it has been demonstrated that at least some of them are strong enough to produce bound states at the lower gap edge. In this case, the sign of $\delta\rho^-$ in this region could easily be opposite to the sign of predominant e-h scattering process since it could have a resonance at negative bias. This appears unlikely, since our investigation of defects in our field of view has found very few pronounced bound state features, and also because a bound state would tend to change the sign of the observed δρ⁻ over a rather narrow range of bias.

c) *Antiphase hole pocket gaps.* Large intensity at small q could be produced by gap structures with sign changes among the hole pockets, or with time reversal symmetry – breaking structures of the s+is type[7] provided the internal phases are not too close to zero. However, one must explain why the sign of $\delta\rho^-$ is opposite that of the e-h processes. This could occur if the impurity potential were nearly diagonal in orbital space, as expected, but had an opposite sign for one orbital channel relative to the others. For example, if the xy pocket $h_3$ had a gap opposite in sign to the inner xz/yz pockets, *and* the defect potential was + for xz/yz and – for xy, scattering between $h_3$ and the xy states on the $h_1$ and $h_2$ pockets would appear as negative (blue) in $\delta\rho^-$. However, there is very little xy weight on $h_1$ and $h_2$. An alternative would be that the off-diagonal elements of the impurity potential might be significant, mixing xy and xz/yz states, and of opposite sign; these off-diagonal potentials have been found to be negligible in microscopic calculations, however[8] effects from changes of hoppings in vicinity of the impurity would need to be considered. Somewhat stronger potentials can enhance this effect, since the impurity T-matrix will generically acquire significant off-diagonal components.

Fig. S1

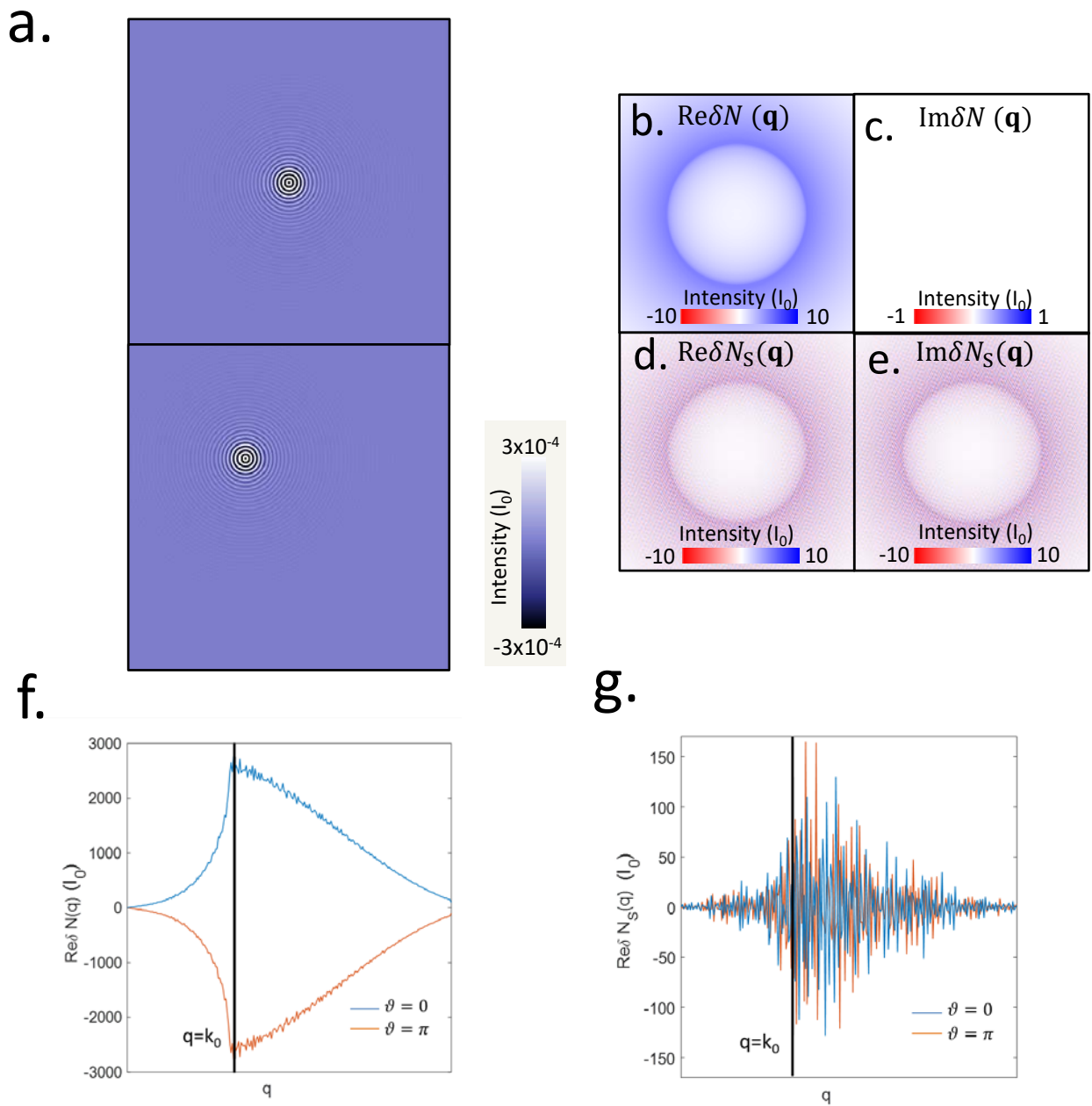

1. **Effect of Shifting the Impurity from the Origin of FOV on The Fourier Transform**
   a. Top: Simulation of Friedel oscillations from an impurity atom located at the origin of the FOV. Bottom: Simulation of Friedel oscillations from an impurity atom shifted from the origin of the FOV. The amplitude units are $I_0$.
   b. Real part of Fourier transform Re $\delta N(\mathbf{q})$ from $\delta N(\mathbf{r})$ from top half of 1a. We use an integer grid; hence the units of Fourier transform are also $I_0$.
   c. Imaginary part of Fourier transform Im $\delta N(\mathbf{q})$ from $\delta N(\mathbf{r})$ from top half of 1a.
   d. Real part of Fourier transform Re $\delta N_S(\mathbf{q})$ from $\delta N_S(\mathbf{r})$ from bottom half of 1a.
   e. Imaginary part of Fourier transform Im $\delta N_S(\mathbf{q})$ from $\delta N_S(\mathbf{r})$ from bottom half of 1a. Rapid oscillations appear in bottom half because impurity is no longer at the origin of the FOV.
   f. Re $\delta N(\mathbf{q})$ from $\delta N(\mathbf{r})$ in 1a for $\vartheta = 0$ and $\vartheta = \pi$, integrated azimuthally from top left in 1b. The signal maintains a definite sign at the peak and flips for the $\pi$ phase shift as expected.
   g. Re $\delta N(\mathbf{q})$ from $\delta N(\mathbf{r})$ in 1a for $\vartheta = 0$ and $\vartheta = \pi$, integrated azimuthally from bottom left in 1b. Rapid oscillations occur because the origin is shifted from the origin of the FOV

Fig. S2

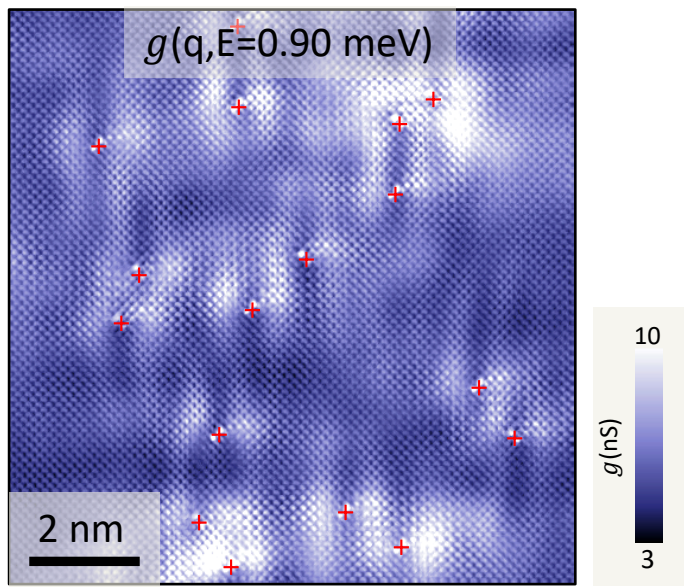
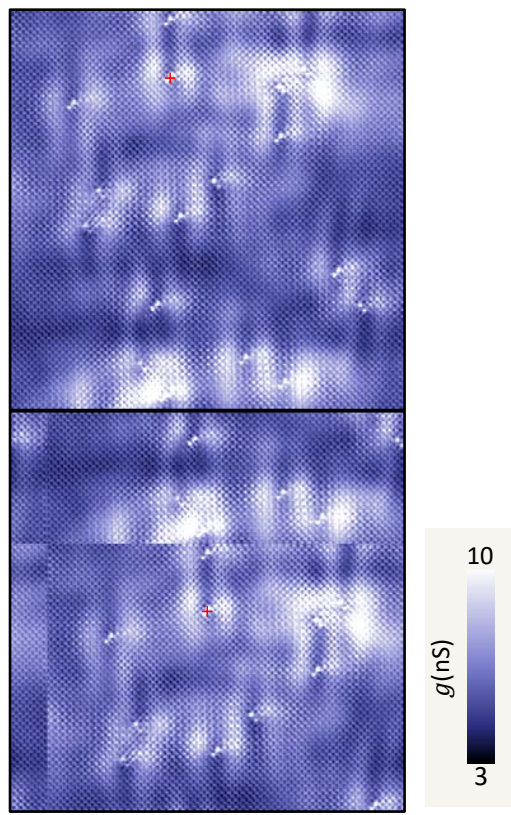
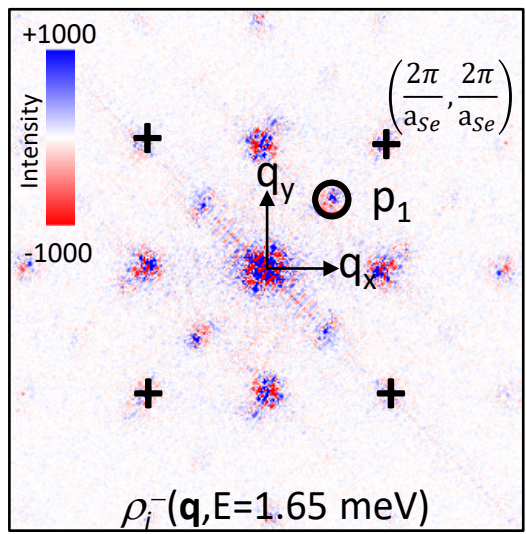
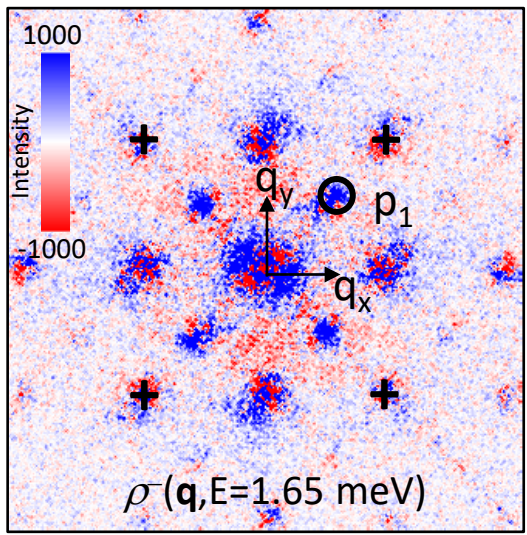

2. **Schematic for MAHAEM Analysis of BQPI in FeSe**

a. FeSe $g(\mathbf{r}, E = 0.9$ meV$)$ layer showing the impurities chosen for the MAHAEM analysis with red crosses. These are missing Fe atoms which are situated between the two bright spots which are due to the LDOS at Se atoms in the top cleaved layer. There are no twin boundaries in this FOV.

b. Top: An impurity (shown by red cross) not at the origin of the FOV. Bottom: Inverse Fourier transform after applying shift theorem to the Fourier transform of a. with the impurity coordinates as shown by the red cross. This impurity has shifted to center now with periodic boundary conditions.

c. $\rho_i^-(\mathbf{q},E=1.65$ meV$)$ calculated using Eq.(13) for the impurity shown in b. Black crosses denote the Bragg peaks. The circle indicates region around the electron-hole scattering vector $p_1$ which is the subject of study.

d. $\rho^-(\mathbf{q},E=1.65$ meV$)$ calculated by summing images like c. for all impurities using Eq.(14). The enhancement in the phase resolved signal can be seen from the intensity of the colors and clear distinction in the red and blue region as the colorscale is quantitatively identical for both 2c and 2d. Black crosses denote the Bragg peaks. The circle indicates region around the electron-hole scattering vector $p_1$ which is the subject of study.

# Fig. S3

a.

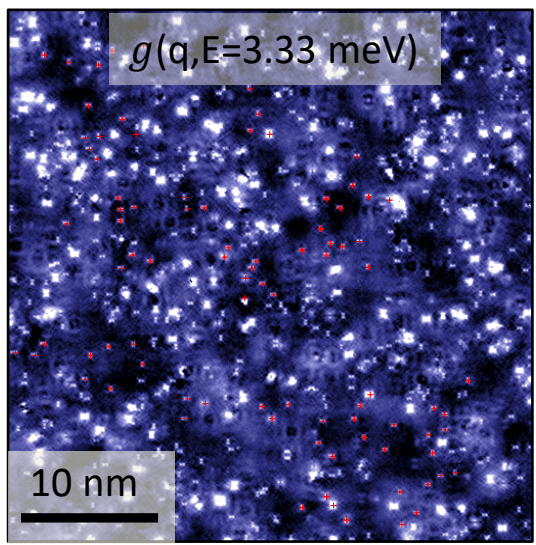

b.

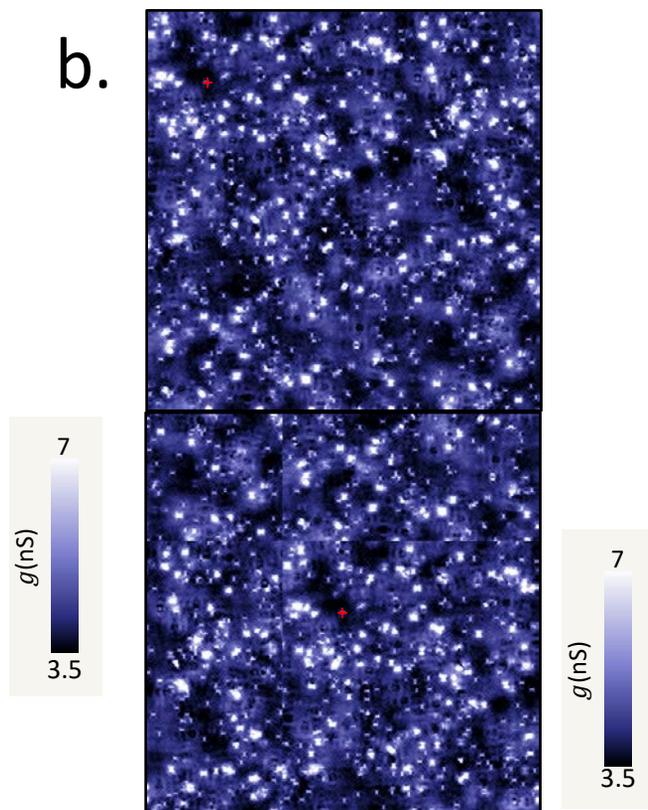

c.

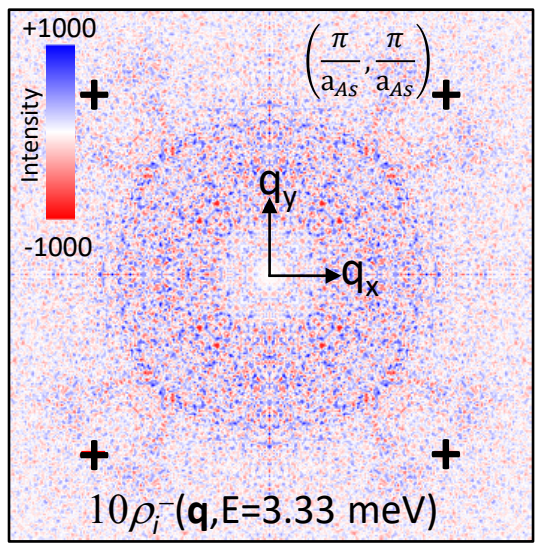

d.

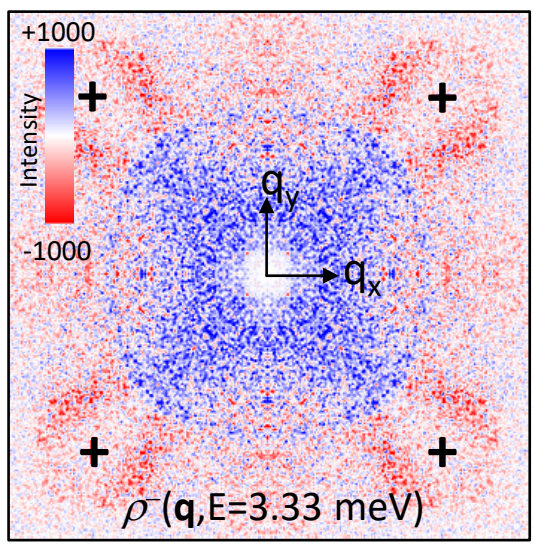

3. **MAHAEM Analysis of electron-hole scattering interference in LiFeAs**

a. LiFeAs $g(\mathbf{r}, E = 3.33 \text{ meV})$ layer showing the impurities chosen for the MAHAEM analysis with red crosses. These are missing Fe vacancies which are situated between the two bright spots which are due to the LDOS at As atoms in the top cleaved layer.

b. Top: An impurity (shown by red cross) not at the origin of the FOV. Bottom: Inverse Fourier transform after applying shift theorem to the Fourier transform of a. with the impurity coordinates as shown by the red cross. This impurity has shifted to center now with periodic boundary conditions.

c. $\rho_i^-(\mathbf{q}, E = 3.33 \text{ meV})$ calculated using Eq.(13) for the impurity shown in b. The intensity is much weaker from a single impurity and the value is enhanced by 10 times. It can also be seen that the phase of features is not clear, there are both red and blue points close to each other.

d. $\rho^-(\mathbf{q}, E = 3.33 \text{ meV})$ calculated by summing images like c. for all impurities using Eq.(14). The enhancement in the phase resolved signal can be seen from the intensity of the colors and clear distinction in the red and blue region as the colorscale is quantitatively identical for both 3c and 3d. A Gaussian mask of $\sigma = 0.68 \text{ Å}^{-1} = 0.8\pi/a$ has been used to remove the bright core emanating from the long-range disorder. In this way, the suppressed area (considering the region suppressed to be a circle of radius half-width-at-half-maximum of this Gaussian) corresponds to ~18% of the total area of the first q-space Brillouin zone The unprocessed data, with no such low-q suppression, is available in Fig. S6a.

Fig. S4

a.
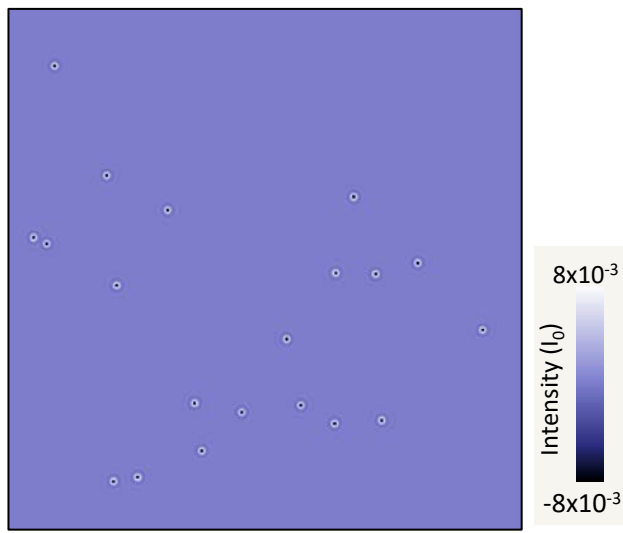

b.
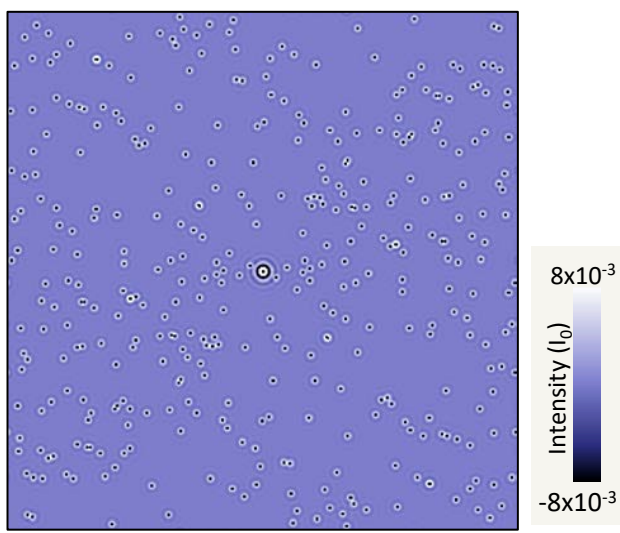

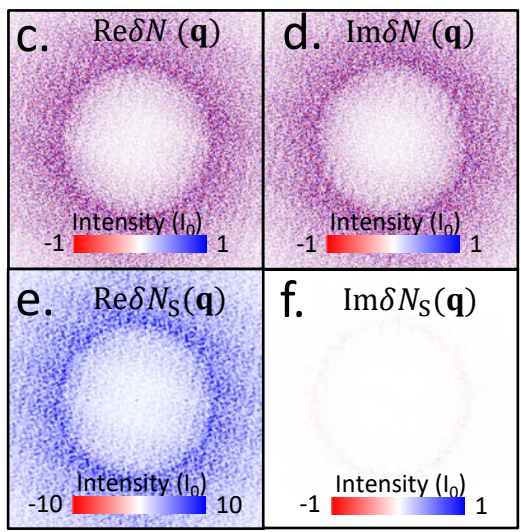

4. **MAHAEM Procedures**
a. Simulation of Friedel oscillations for 20 impurities in a FOV. The amplitude units are $I_0$.
b. Result of addition of images where each impurity is shifted to the center as described in supplementary note 3. The scattering from the impurity at the center is much enhanced, while other impurities become weaker due to averaging out.
c. Real part of the Fourier transform of the image in a. Since an integer grid is used, the units are $I_0$.
d. Imaginary part of the Fourier transform of the image in a.
e. Real part of the Fourier transform of the image in b. Notice the color scale. It is much enhanced in intensity, demonstrating the usefulness of our multi-atom procedure.
f. Imaginary part of the Fourier transform of the image in b. It is much smaller as compared to the real part as seen in e.

Fig. S5

a.
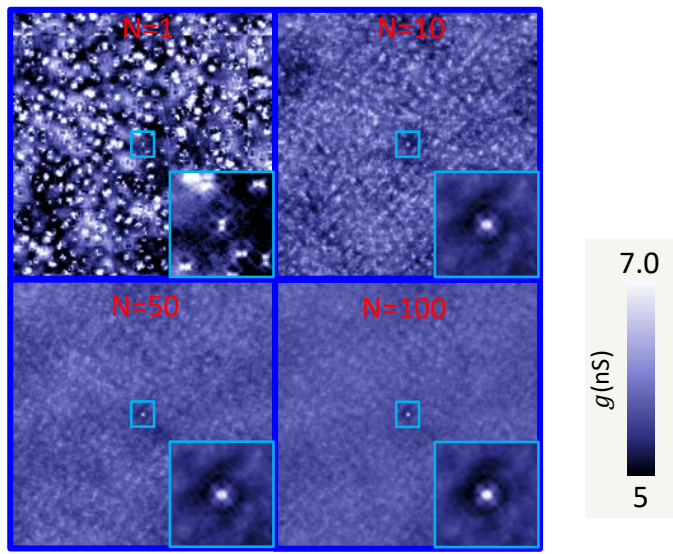

b.
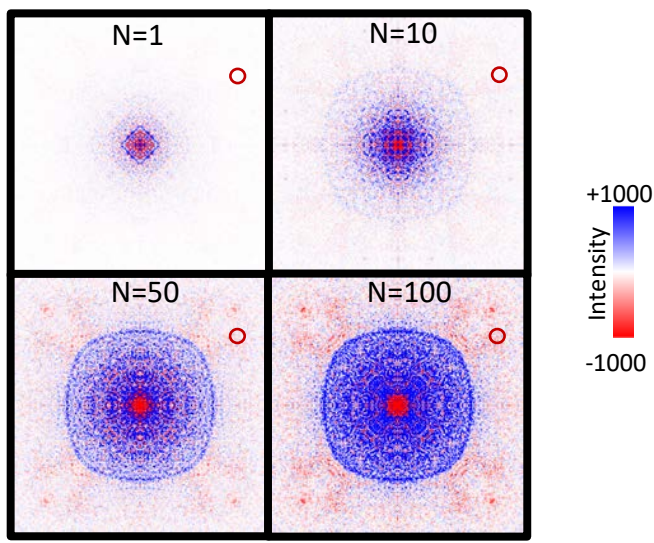

c.
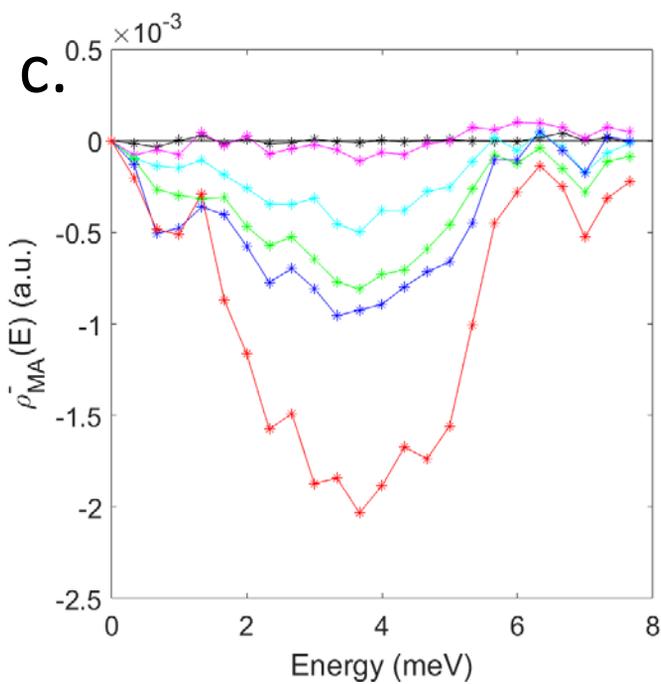

5. **Example of MAHAEM procedures with varying N**
   a. Real-space conductance maps $\sum_{i=1}^{N} g^S(\mathbf{r}, E = 5.33 \text{ meV})$ calculated using Eq.(10) for N=1, 10, 50 and 100 real scatterers. The inset shows the zoomed-in version of the area surrounding the central impurity (light blue square box).
   b. Symmetrized $\bar{\rho}_{MA}(\mathbf{q}, E) = \sum_{i=1}^{N} \bar{\rho_i}(\mathbf{q}, E)$ for $E = 5.33$ meV calculated using Eq.(14) for N=1, 10, 50 and 100 real scatterers.
   c. The $\bar{\rho}_{MA}(E = 5.33 \text{ meV})$ integrated over the circular region as shown in b. for N=1, 10, 50 and 100 real scatterers.

Fig. S6

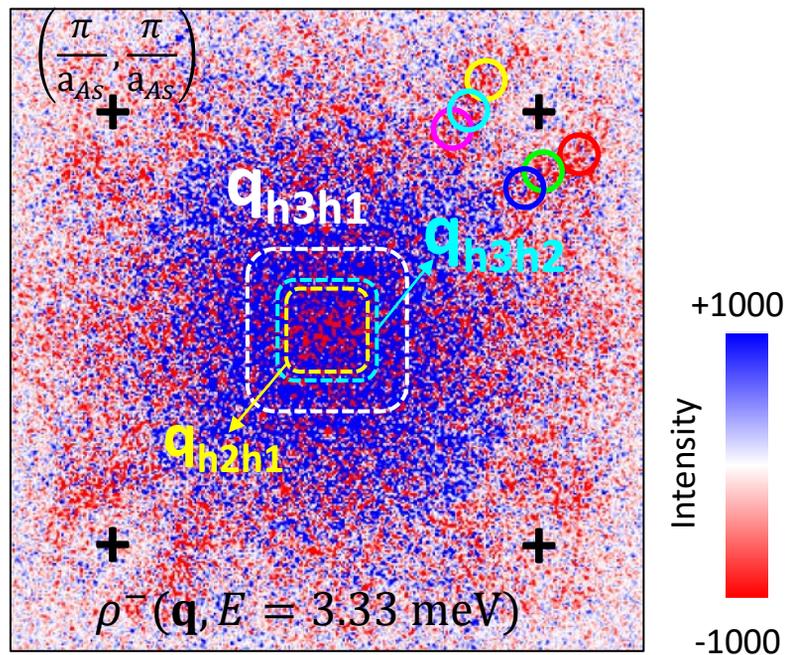

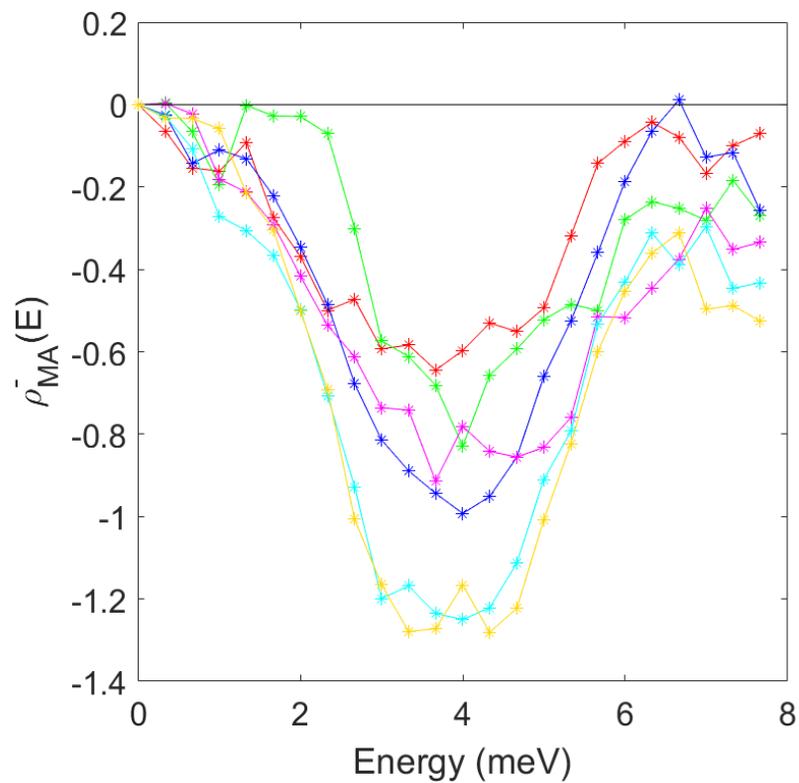

6. **Robustness of MAHAEM in LiFeAs**

a. Unprocessed $\rho^-_{MA}(\mathbf{q}, E = 3.33 \text{ meV})$ showing circles of the same size as used in Fig. 4c. on different locations on the horn shaped feature which denotes the scattering from small hole pockets around Γ- point to electron pockets around X-points. The dashed contours are overlaid to indicate inter-hole scatterings $q_{h_3h_1}$(white), $q_{h_3h_2}$(cyan) and $q_{h_2h_1}$(yellow) (from largest to smallest) estimated from experimental QPI in Fig. 2D of Ref. 9.

b. Measured $\rho^-_{MA}(E)$ integrated over the circles with colors corresponding to image in a. The $\rho^-_{MA}(E)$ is quite similar for all these locations and does not change the sign and the negative values peak at $E \approx \sqrt{\Delta_1\Delta_2}$. This is as expected from the detailed theoretical calculations based on experimental values as shown in Fig. 3d and Fig. 4c in the main text.